\documentclass{svjour3}

\usepackage{cite}

\usepackage{graphicx}

\usepackage{dcolumn}

\journalname{Journal of Mathematical Chemistry}

\begin{document}

\title{Particle correlation from uncorrelated non Born-Oppenheimer SCF wavefunctions}

\author{Paolo Amore \and Francisco M Fern\'andez}

\institute{Paolo Amore \at Facultad de Ciencias, Universidad de
Colima, Bernal D\'iaz del Castillo 340, Colima, Colima, Mexico
\email{paolo.amore@gmail.com} \and Francisco M Fern\'andez \at
INIFTA (UNLP, CCT La Plata-CONICET), Divisi\'on Qu\'imica
Te\'orica, Blvd. 113 S/N, Sucursal 4, Casilla de Correo 16, 1900
La Plata, Argentina \email{fernande@quimica.unlp.edu.ar}}

\date{Received: date / Accepted: date}

\maketitle

\begin{abstract}
We analyse a nonadiabatic self-consistent field method by means of
an exactly-solvable model. The method is based on nuclear and
electronic orbitals that are functions of the cartesian
coordinates in the laboratory-fixed frame. The kinetic energy of
the center of mass is subtracted from the molecular Hamiltonian
operator in the variational process. The results for the simple
model are remarkably accurate and show that the integration over
the redundant cartesian coordinates leads to couplings among the
internal ones.


\end{abstract}

\section{Introduction}

\label{sec:intro}

Typical quantum-mechanical treatments of molecular systems are
based on the Born-Oppenheimer (BO) approximation that separates
the motions of nuclei and electrons\cite{BJ00}. One first solves
an eigenvalue equation for the electrons in the field of nuclei
clamped at some predetermined points in space and thus builds the
potential-energy surface (PES). Then one solves an equation for
the nuclei moving on that PES and obtains the allowed energies of
the molecule. In many cases the calculation is restricted to the
neighborhood of the minimum of the PES in order to determine the
molecular geometry\cite{SO96}. Accurate PES's are useful for
chemical-kinetic studies\cite{F73}.

There has recently been great interest on the calculation of
molecular properties by means of nonadiabatic approaches; i.e.
without resorting to the BO approximation. Since the
eigenfunctions of the Hamiltonian operator of the molecule
$\hat{H}$ are not square integrable with respect to the $3N$
coordinates that describe the $N$ particles (electrons plus
nuclei) in the laboratory-fixed coordinate axes one has to remove
the unbounded motion of the molecular center of mass and the three
corresponding coordinates. Thus one is left with the
Schr\"{o}dinger equation for the resulting molecular Hamiltonian
operator in the molecule-fixed coordinate axes $H_{M}$ that
depends on $3N-3$ coordinates and momenta. For brevity, from now
on we call those coordinates absolute and relative (or internal),
respectively. By means of the variational method one obtains an
approximate eigenfunction of $H_{M}$ that should be also
eigenfunction of the operators that commute with $H_{M}$, such as,
for example, spin, angular momentum operator in relative
coordinates, etc\cite {W91,FE10}.

There are several strategies for solving the Schr\"{o}dinger
equation without the BO approximation. One of them is based on
explicitly correlated Gaussian functions of the relative
coordinates of all the electrons and nuclei in the molecule (for a
comprehensive review see Bubin et al\cite{BCA05}). The trial
function constructed from those Gaussians satisfies the
permutational symmetry of identical particles and contains several
parameters that are to be determined according to the variational
method. If the Gaussians are located at one nuclei it is not
difficult to choose the variational wavefunction to be
eigenfunction of the angular-momentum operator. In some cases it
is convenient to place the centers of the Gaussians on different
space points and it is more difficult to force the variational
function to be eigenfunction of the angular-momentum operator. In
these calculations the authors explicitly expressed the
variational function and the Hamiltonian operator in relative
coordinates\cite{BCA05} (and references therein).

Another widely spread strategy is based on uncorrelated functions
and the SCF approach. In this case the variational function is
written as a product of one-particle functions with the
appropriate permutational symmetry. Since there are nuclear
orbitals in addition to electronic ones this approach has been
termed nuclear orbital plus molecular orbital (NOMO)
theory\cite{TMNI98,N02,NHM05, HN06,MHN06,N07} and also (but no
longer in use now) dynamic extended molecular orbital (DEMO)
method\cite{TO00}. Since such orbitals are expressed in terms of
the $3N$ absolute cartesian coordinates one has to be careful to
avoid the spurious contribution of the kinetic energy of the
center of mass\cite{F10}. In order to avoid this problem Nakai et
al\cite{NHM05, HN06,MHN06,N07} proposed to subtract the kinetic
energy of the molecular center of mass from the Hamiltonian
operator during the variational optimization of the trial
wavefunction and developed the translation-free NOMO (TF-NOMO).

Although the NOMO approach is a nonadiabatic method its
implementation is reminiscent of the BO approximation in that the
orbitals are located at the ``experimental
geometries''\cite{NHM05}. Note that such a concept is rather alien
to the nonadiabatic quantum-mechanical calculation just outlined
(compare it with the more rigorous approach described by Bubin et
al\cite{BCA05}). The properly symmetrized product of orbitals
located at different points and expressed in terms of absolute
coordinates is not an eigenfunction of the angular-momentum
operator\cite{NHM05, HN06,MHN06,N07}. For that reason several
rotational states are expected to contribute to the optimized
variational function.

In order to remove the contribution of rotational states with
angular-momentum quantum number $J>0$ Nakai et al\cite{NHM05,
HN06,MHN06,N07} proposed a rotation-free NOMO that consists of
subtracting also the rotational kinetic energy from the total
Hamiltonian operator. The resulting approach is named translation-
and rotation-free NOMO\ (TRF-NOMO). However, the removal of the
spurious rotational kinetic energy in this way is not exact as in
the case of the translation kinetic energy as argued by
Sutcliffe\cite{S05}.

The purpose of this paper is the analysis of the performance of
the TF-NOMO and the effect  of using  absolute cartesian
coordinates in the trial wavefunction. Since the treatment of
realistic examples may be rather cumbersome we apply the method to
a simple exactly solvable model.

In order to make this paper sufficiently self-contained, and to
facilitate the discussion throughout, in Sec.~\ref{sec:system} we
outline the separation of the kinetic energy of the center of mass
and the construction of the molecular Hamiltonian in relative
coordinates. In Sec.~\ref{sec:model} we apply the TF-NOMO to an
exactly solvable model in order to test the accuracy of this
approach and the effect of using the absolute coordinates instead
of the internal ones. Finally, in Sec.~\ref{sec:conclusions} we
discuss the results and draw conclusions.

\section{Molecular Hamiltonian}

\label{sec:system}

In this section we outline some general properties of the
nonrelativistic Hamiltonian operator for a system of $N$ charged
point particles with only Coulomb interactions. The results are
well known and have been discussed by several authors in different
contexts (see, for example, the review by Fern\'{a}ndez and
Echave\cite{FE10} and the references therein). The nonrelativistic
Hamiltonian operator for a molecule can be written as
\begin{eqnarray}
\hat{H} &=&\hat{T}+\hat{V},  \nonumber \\
\hat{T} &=&\sum_{i=1}^{N}\frac{\hat{p}_{i}^{2}}{2m_{i}},  \nonumber \\
V &=&\frac{1}{4\pi \epsilon _{0}}\sum_{i=1}^{N-1}\sum_{j=i+1}^{N}\frac{%
Q_{i}Q_{j}}{r_{ij}}  \label{eq:H=T+V_full}
\end{eqnarray}
where $m_{i}$ is the mass of particle $i$, $Q_{i}=-e$ or $%
Q_{i}=Z_{i}e$ are the charges of either an electron or nucleus,
respectively, and $r_{ij}=|\mathbf{r}_{i}-\mathbf{r}_{j}|$ is the
distance between particles $i$ and $j$ located at the points
$\mathbf{r}_{i}$ and $\mathbf{r}_{j}$, respectively, from the
origin of the laboratory-fixed coordinate system. In the
coordinate representation $\mathbf{\hat{p}}_{i}=-i\hbar \nabla
_{i}$.

Since the the uniform translation of all the particles
$\hat{U}(\mathbf{a})\mathbf{r}_{i}\hat{U}(\mathbf{a})^{\dagger
}=\mathbf{r}_{i}+\mathbf{a}$ leaves the Coulomb potential
invariant $\hat{U}(\mathbf{a})V\hat{U}(\mathbf{a})^{\dagger }=V$,
then the eigenfunctions of the translation--invariant Hamiltonian
operator (\ref {eq:H=T+V_full}) are not square integrable. For
that reason we separate the motion of the center of mass and
define translation-invariant internal coordinates by means of a
linear transformation
\begin{equation}
\mathbf{r}_{j}^{\prime }=\sum_{i}t_{ji}\mathbf{r}_{i}  \label{eq:r'->r}
\end{equation}
It is our purpose to keep the transformation (\ref{eq:r'->r}) as general as
possible so that it applies to a wide variety of nonadiabatic approaches. We
arbitrarily choose $\mathbf{r}_{1}^{\prime }$ to be the coordinate of the
center of mass
\begin{equation}
t_{1i}=\frac{m_{i}}{M},\;M=\sum_{i}m_{i}  \label{eq:t1i}
\end{equation}
and $\mathbf{r}_{j}^{\prime }$, $j>1$ the translational-invariant
coordinates
\begin{equation}
\sum_{i}t_{ji}=0,\;j>1  \label{eq:t_CM_2}
\end{equation}
Note that if the coefficients of the linear transformation (\ref{eq:r'->r})
satisfy equations (\ref{eq:t1i}) and (\ref{eq:t_CM_2}) then
\begin{equation}
\hat{U}(\mathbf{a})\mathbf{r}_{j}^{\prime }\hat{U}(\mathbf{a})^{\dagger }=%
\mathbf{r}_{j}^{\prime }+\mathbf{a}\delta _{j1}  \label{eq:Ur'U+}
\end{equation}
The choice of the coefficients of the transformation (\ref{eq:r'->r}) for
the translational-invariant variables $\mathbf{r}_{j}^{\prime }$, $j>1$ is
arbitrary as long as they satisfy Eq.~(\ref{eq:t_CM_2}) (for a more detailed
discussion see \cite{FE10}).

As a result of the change of variables, the total Hamiltonian operator reads

\begin{eqnarray}
\hat{H} &=&-\frac{\hbar ^{2}}{2M}\nabla _{1}^{\prime 2}+\hat{H}_{M}
\nonumber \\
\hat{H}_{M} &=&-\frac{\hbar ^{2}}{2}\sum_{j>1}\sum_{k>1}\left( \sum_{i}\frac{%
t_{ji}t_{ki}}{m_{i}}\right) \nabla _{j}^{\prime }\nabla _{k}^{\prime }+\frac{%
1}{4\pi \epsilon _{0}}\sum_{i=1}^{N-1}\sum_{j=i+1}^{N}\frac{Q_{i}Q_{j}}{%
r_{ij}}  \label{eq:H_M}
\end{eqnarray}
where $\hat{H}_{M}$ is the internal or molecular Hamiltonian
operator. The explicit form of the interparticle distances
$r_{ij}$ in terms of the new coordinates $r_{k}^{\prime }$ may be
rather cumbersome in the general case but there are particular
choices that are suitable for the calculation of the integrals
necessary for the application of the variational method\cite
{FE10,BCA05}. The treatment of the simple model in
Sec.~\ref{sec:model} shows one of those particular
transformations.

For future reference it is convenient to define the center of mass
and relative kinetic-energy operators
\begin{eqnarray}
\hat{T}_{CM} &=&-\frac{\hbar ^{2}}{2M}\nabla _{1}^{\prime 2}
\label{eq:TCM_gen_1} \\
\hat{T}_{rel} &=&-\frac{\hbar ^{2}}{2}\sum_{j>1}\sum_{k>1}\left( \sum_{i}%
\frac{t_{ji}t_{ki}}{m_{i}}\right) \nabla _{j}^{\prime }\nabla _{k}^{\prime }
\label{eq:Trel}
\end{eqnarray}
respectively, so that $\hat{T}=\hat{T}_{CM}+\hat{T}_{rel}$,
$\hat{H}_{M}=\hat{T}_{rel}+V$, and
$\hat{H}=\hat{T}_{CM}+\hat{H}_{M}$.

The inverse transformation $\mathbf{t}^{-1}$ exists and gives us the old
coordinates in terms of the new ones:
\begin{equation}
\mathbf{r}_{i}=\sum_{j}\left( \mathbf{t}^{-1}\right) _{ij}\mathbf{r}%
_{j}^{\prime }  \label{eq:r->r'}
\end{equation}
According to equations (\ref{eq:Ur'U+}) and (\ref{eq:r->r'}) we have $\hat{U%
}(\mathbf{a})\mathbf{r}_{i}\hat{U}(\mathbf{a})^{\dagger }=\left( \mathbf{t}%
^{-1}\right) _{i1}\mathbf{a}+\mathbf{r}_{i}$ from which we conclude that
\begin{equation}
\left( \mathbf{t}^{-1}\right) _{i1}=1,\;i=1,2,\ldots ,N  \label{eq:(1/t)i1}
\end{equation}
In order to understand the meaning of this result note that the momentum
conjugate to $\mathbf{r}_{i}^{\prime }$ is given by the transformation
\begin{equation}
\mathbf{\hat{p}}_{i}^{\prime }=\sum_{j}\left( \mathbf{t}^{-1}\right) _{ji}%
\mathbf{\hat{p}}_{j}  \label{eq:p'->p}
\end{equation}
so that the linear momentum of the center of mass
\begin{equation}
\mathbf{\hat{p}}_{1}^{\prime }=\sum_{j}\mathbf{\hat{p}}_{j}  \label{eq:p'_1}
\end{equation}
is precisely the total linear momentum of the molecule. We also appreciate
that
\begin{equation}
T_{CM}=\frac{\mathbf{\hat{p}}_{1}^{\prime }\cdot \mathbf{\hat{p}}%
_{1}^{\prime }}{2M}=\frac{1}{2M}\sum_{i=1}^{N}\sum_{j=1}^{N}\mathbf{\hat{p}}%
_{i}\cdot \mathbf{\hat{p}}_{j}  \label{eq:TCM_gen_2}
\end{equation}

The eigenfunctions of the total Hamiltonian operator (\ref{eq:H=T+V_full})
are of the form
\begin{equation}
\Psi (\mathbf{r}_{1},\ldots ,\mathbf{r}_{N})=e^{i\mathbf{k}\cdot \mathbf{r}%
_{1}^{\prime }}\psi (\mathbf{r}_{2}^{\prime },\ldots ,\mathbf{r}_{N}^{\prime
})  \label{eq:Psi_tot_gen}
\end{equation}
with the appropriate permutational symmetry for the electrons and
nuclei. We are of course interested in the wavefunction for the
internal degrees of freedom $\psi (\mathbf{r}_{2}^{\prime },\ldots
,\mathbf{r}_{N}^{\prime })$ that provides the relevant molecular
properties. For this reason we should use a trial function of the
corresponding coordinates $\varphi (\mathbf{r}_{2}^{\prime
},\ldots ,\mathbf{r}_{N}^{\prime })$ and apply the variational
method with the relative or molecular Hamiltonian operator
$\hat{H}_{M}$\cite {BCA05}.

Nakai et al\cite{NHM05, HN06,MHN06,N07} proposed an alternative
route based on a trial function of the absolute cartesian
coordinates $\varphi (\mathbf{r}_{1},\ldots ,\mathbf{r}_{N})$ and
applied the variational method to
\begin{equation}
W=\frac{\left\langle \varphi \right| \hat{H}-\hat{T}_{CM}\left| \varphi
\right\rangle }{\left\langle \varphi \right| \left. \varphi \right\rangle }
\label{eq:W_Nak_gen}
\end{equation}
More precisely, they resorted to the Hartree-Fock method with
electronic and nuclear orbitals. In this equation one integrates
with respect to $d\mathbf{r}_{1}d\mathbf{r}_{2}\ldots
d\mathbf{r}_{N}$ (including spin if necessary) and the trial
function $\varphi (\mathbf{r}_{1},\ldots ,\mathbf{r}_{N})$ is
square integrable with respect to all those $3N$ electronic and
nuclear variables. This approach is called TF-NOMO method and is
an improvement over the translation-contaminated NOMO TC-NOMO
method (like, for example, the DEMO\cite{TO00}) that is based on
the variational method for $\hat{H}$. Note that the domain of the
NOMO trial function $\varphi (\mathbf{r}_{1},\ldots
,\mathbf{r}_{N})$ is $R^{3N}$ and that for the molecular
wavefunction $\psi (\mathbf{r}_{2}^{\prime },\ldots
,\mathbf{r}_{N}^{\prime })$ is $R^{3N-3}$. Thus, from a
quantum-mechanical point of view they belong to different state
spaces.

If we rewrite the arguments of the trial function in terms of
internal coordinates $\varphi (\mathbf{r}_{1},\ldots
,\mathbf{r}_{N})=\tilde{\varphi}(\mathbf{r}_{1},\mathbf{r}_{2}^{\prime
},\ldots ,\mathbf{r}_{N}^{\prime })$ then we appreciate that the
probability distribution of the internal variables
\begin{equation}
\rho (\mathbf{r}_{2}^{\prime },\ldots ,\mathbf{r}_{N}^{\prime
})=\int \tilde{\varphi}(\mathbf{r}_{1},\mathbf{r}_{2}^{\prime
},\ldots ,\mathbf{r}_{N}^{\prime })^{2}\,d\mathbf{r}_{1}
\label{eq:rho_gen}
\end{equation}
may exhibit some degree of correlation even though the trial
function $\varphi (\mathbf{r}_{1},\ldots ,\mathbf{r}_{N})$ is
merely a product of nuclear and electronic orbitals with the
appropriate permutational symmetry\cite{NHM05, HN06,MHN06,N07}.
The analysis of the effect of this particle correlation in a
realistic molecular system appears to be rather complicated; for
this reason in what follows we resort to a quite simple example.

\section{Simple model}

\label{sec:model}

In order to have a clearer understanding of the TF-NOMO
method\cite{NHM05, HN06,MHN06,N07} we apply it to a simple
exactly-solvable model. We are only interested in the removal of
the translational contamination because getting rid of the
rotational one does not appear to be so simple\cite{S05}.
Therefore, a one--dimensional model with at least three particles
and a translation-invariant potential will suffice. Our model
consists of three particles of masses $M_{1}$, $M_{2}$ and $M_{3}$
that move in one dimension and interact through forces that follow
Hooke's law:
\begin{eqnarray}
\hat{H} &=&-\frac{\hbar ^{2}}{2M_{1}}\frac{\partial ^{2}}{\partial X_{1}^{2}}%
-\frac{\hbar ^{2}}{2M_{2}}\frac{\partial ^{2}}{\partial X_{2}^{2}}-\frac{%
\hbar ^{2}}{2M_{3}}\frac{\partial ^{2}}{\partial X_{3}^{2}} \\
&&+\frac{1}{2}\left[ K_{12}\left( X_{1}-X_{2}\right) ^{2}+K_{13}\left(
X_{1}-X_{3}\right) ^{2}+K_{23}\left( X_{2}-X_{3}\right) ^{2}\right]
\label{eq:Ham_X}
\end{eqnarray}
where $K_{ij}$ are the force constants.

In order to reduce the number of parameters to a minimum we first
assume that the three particles are identical, so that
$M_{1}=M_{2}=M_{3}=M$ and $K_{12}=K_{13}=K_{23}=K$. We define
dimensionless coordinates $x_{i}=X_{i}/L$, where $L=\hbar
/(m\omega )$ and $\omega =\sqrt{K/M}$ and obtain the dimensionless
Hamiltonian
\begin{eqnarray}
\hat{H}_{d} &=&\frac{\hat{H}}{\hbar \omega }=-\frac{1}{2}\left( \frac{%
\partial ^{2}}{\partial x_{1}^{2}}+\frac{\partial ^{2}}{\partial x_{2}^{2}}+%
\frac{\partial ^{2}}{\partial x_{3}^{2}}\right)  \nonumber \\
&&+\frac{1}{2}\left[ \left( x_{1}-x_{2}\right) ^{2}+\left(
x_{1}-x_{3}\right) ^{2}+\left( x_{2}-x_{3}\right) ^{2}\right]
\label{eq:Ham_x}
\end{eqnarray}
and the dimensionless energy $\epsilon =E/(\hbar \omega )$.

We separate the motion of the center of mass by means of the transformation
\begin{eqnarray}
q_{1} &=&\frac{1}{3}\left( x_{1}+x_{2}+x_{3}\right)  \nonumber \\
q_{2} &=&x_{2}-x_{1}  \nonumber \\
q_{3} &=&x_{3}-x_{1}  \label{eq:q(x)}
\end{eqnarray}
where $q_{1}$ is the coordinate of the center of mass and $q_{2}$
and $q_{3}$ are the coordinates of particles 2 and 3 with respect
to the coordinate origin located arbitrarily at particle 1. These
$q_{j}$'s are the $\mathbf{r}_{j}^{\prime }$'s of
Sec.~\ref{sec:system} and the transformation (\ref{eq:q(x)})
satisfies the equations discussed there. We thus obtain
\begin{equation}
\hat{H}_{d}=\hat{T}_{CM}+\hat{H}_{r}  \label{eq:Ham_q}
\end{equation}
that is the sum of the dimensionless kinetic energy of the center
of mass $\hat{T}_{CM}$ and the dimensionless Hamiltonian operator
for the relative motion $\hat{H}_{r}$ ($\hat{H}_{M}$ in the
general discussion of Sec.~\ref{sec:system})
\begin{eqnarray}
\hat{T}_{CM} &=&-\frac{1}{6}\frac{\partial ^{2}}{\partial q_{1}^{2}}
\nonumber \\
\hat{H}_{r} &=&-\frac{\partial ^{2}}{\partial q_{2}^{2}}-\frac{\partial ^{2}%
}{\partial q_{3}^{2}}-\frac{\partial ^{2}}{\partial q_{2}\partial q_{3}}%
+q_{2}^{2}+q_{3}^{2}-q_{2}q_{3}  \label{eq:TCM,Hr}
\end{eqnarray}
The eigenfunctions are of the form
\begin{equation}
\Psi (q_{1},q_{2},q_{3})=e^{ikq_{1}}\psi (q_{1},q_{2})  \label{eq:Phi}
\end{equation}
where $\psi (q_{1},q_{2})$ is an eigenfunction of $\hat{H}_{r}$
and $-\infty <k <\infty$. Note that $\Psi (q_{1},q_{2},q_{3})$ is
not square integrable with respect to $dx_{1}dx_{2}dx_{3}$ as
expected from the fact that the motion of the center of mass is
unbounded. The total dimensionless energy is $\epsilon
_{T}=\epsilon +k^{2}/6$, where $\epsilon $ is the dimensionless
energy of the relative motion (an eigenvalue of $\hat{H}_{r}$).

If we try the correlated gaussian function\newline $\varphi
(q_{2},q_{3})=\left[ \left( 4a^{2}-b^{2}\right) /\pi ^{2}\right]
^{1/4}\exp \left[ -a\left( q_{2}^{2}+q_{3}^{2}\right)
-bq_{2}q_{3}\right] $,\newline where $4a^{2}-b^{2}>0$, then we
obtain the exact ground-state eigenfunction of $\hat{H}_{r}$
\begin{equation}
\psi _{00}(q_{2},q_{3})=\frac{1}{\sqrt{\pi }}\exp \left[
-\frac{\sqrt{3}}{3}\left( q_{2}^{2}-q_{2}q_{3}+q_{3}^{2}\right)
\right]  \label{eq:phi_exact}
\end{equation}
with dimensionless energy $\epsilon _{00}=\sqrt{3}$.

As uncorrelated variational function in the absolute coordinates we try
\begin{equation}
\varphi (x_{1},x_{2},x_{3})=\left( \frac{2a}{\pi }\right) ^{3/4}\exp \left[
-a\left( x_{1}^{2}+x_{2}^{2}+x_{3}^{2}\right) \right]   \label{eq:phi_var}
\end{equation}
that is square integrable with respect to $dx_{1}dx_{2}dx_{3}$.
Note that there is a redundant coordinate because we need just two
variables to describe the bound states of this model as shown in
Eq.~(\ref{eq:phi_exact}). This trial function is our simple
version of a NOMO one. The optimal value of the variational
parameter $a$ is determined by the minimum of
\begin{equation}
W(a)=\left\langle \hat{H}-\hat{T}_{CM}\right\rangle   \label{eq:W(a)}
\end{equation}
as proposed by Nakai et al\cite{NHM05, HN06,MHN06,N07}, where
\begin{equation}
\hat{T}_{CM}=-\frac{1}{6}\left( \frac{\partial ^{2}}{\partial x_{1}^{2}}+%
\frac{\partial ^{2}}{\partial x_{2}^{2}}+\frac{\partial ^{2}}{\partial
x_{3}^{2}}+2\frac{\partial ^{2}}{\partial x_{1}\partial x_{2}}+2\frac{%
\partial ^{2}}{\partial x_{1}\partial x_{3}}+2\frac{\partial ^{2}}{\partial
x_{2}\partial x_{3}}\right)   \label{eq:TCM_mod}
\end{equation}
is the kinetic-energy operator for the center of mass in the
absolute coordinates. Note that present $\hat{T}_{CM}$ is what
Nakai et al\cite{NHM05} call $\hat{T}_{T}$ and approximate by
$\hat{T}_{n,T}$ in their calculations, and present approach is the
straightforward application of the TF-NOMO to a simple
one-dimensional model.

At first sight it is surprising that the variational function with
the optimal value $a=\sqrt{3}/2$ of the adjustable parameter
yields the exact energy $W(\sqrt{3}/2)=\sqrt{3}$. However, it is
not the only striking fact because the expectation values of any
function of $x_{2}-x_{1}$ and $x_{3}-x_{1}$ (like, for example,
$(x_{2}-x_{1})^{2}$, $(x_{2}-x_{1})^{4}$,
$(x_{2}-x_{1})(x_{3}-x_{1})$, etc) are exact too. In spite of this
remarkable agreement the exact and approximate wavefunctions are
not the same as follows from the fact that $\hat{T}_{CM}\psi
_{00}=0$ and $\left\langle \varphi |\hat{T}_{CM}|\varphi
\right\rangle =\sqrt{3}/4$. We will explain these curious results
later on; for the time being note that the results of Nakai et
al\cite{NHM05} for $E_{tot}^{TRC}-E_{tot}^{TF}=\left\langle \Phi
_{0}\right| \hat{T}_{n,T}\left| \Phi _{0}\right\rangle \approx
\left\langle \Phi _{0}\right| \hat{T}_{T}\left| \Phi
_{0}\right\rangle $ yield the spurious contribution of the kinetic
energy of the center of mass to the molecular energy. They state
that this problem is due to the fact that the Gaussian functions
are unsuitable for describing the translational energy but it is
clear that a set of Gaussian functions in relative coordinates
will not exhibit such undesirable behavior. In other words, the
translational contamination is a consequence of adopting the
absolute coordinates and not a result of the choice of Gaussian
functions. The same argument applies to the rotational
contamination; Bubin et al\cite{BCA05} show how to obtain Gaussian
states with zero angular-momentum quantum number ($J=0$).

We may suspect that the unexpected success of the variational
approach is partly due to the symmetry of the problem (three
identical particles). In order to break it with the slightest
modification of our model we choose $K_{12}=K_{13}=K\neq K_{23}$
and define $\lambda =K_{23}/K$, so that
\begin{eqnarray}
\hat{H}_{d} &=&-\frac{1}{2}\left( \frac{\partial ^{2}}{\partial x_{1}^{2}}+%
\frac{\partial ^{2}}{\partial x_{2}^{2}}+\frac{\partial ^{2}}{\partial
x_{3}^{2}}\right)  \nonumber \\
&&+\frac{1}{2}\left[ \left( x_{1}-x_{2}\right) ^{2}+\left(
x_{1}-x_{3}\right) ^{2}+\lambda \left( x_{2}-x_{3}\right) ^{2}\right]
\label{eq:Ham_x_2}
\end{eqnarray}
and
\begin{equation}
\hat{H}_{r}=-\frac{\partial ^{2}}{\partial q_{2}^{2}}-\frac{\partial ^{2}}{%
\partial q_{3}^{2}}-\frac{\partial ^{2}}{\partial q_{2}\partial q_{3}}+\frac{%
1}{2}\left[ q_{2}^{2}+q_{3}^{2}+\lambda (q_{2}-q_{3})^{2}\right]
\label{eq:Hr_2}
\end{equation}
Since the masses remain the same the transformation from absolute
to relative coordinates and the form of $T_{CM}$ are still given
by equations (\ref{eq:q(x)}) and (\ref{eq:TCM_mod}), respectively.

The exact ground-state wavefunction and energy are given by
\begin{eqnarray}
\psi _{00}(q_{2},q_{3}) &=&\frac{\left( 4a^{2}-b^{2}\right) ^{1/4}}{\sqrt{%
\pi }}\exp \left[ -a\left( q_{2}^{2}+q_{3}^{2}\right) +bq_{2}q_{3}\right]
\nonumber \\
a &=&\frac{\sqrt{6}\left( \sqrt{6\lambda +3}+3\lambda +2\right) \sqrt{%
3\lambda +2-\sqrt{6\lambda +3}}}{12\left( 3\lambda +1\right) }  \nonumber \\
b &=&\frac{\sqrt{6}\sqrt{3\lambda +2-\sqrt{6\lambda +3}}}{6}
\label{eq:phi_exact_lam}
\end{eqnarray}
and

\begin{equation}
\epsilon _{00}=\frac{\sqrt{2}\left( 2\sqrt{2\lambda +1}+\sqrt{3}\left(
\lambda +1\right) \right) \sqrt{3\lambda +2-\sqrt{6\lambda +3}}}{2\left(
3\lambda +1\right) }  \label{eq:e_exact_lam}
\end{equation}
respectively.

As in the preceding example we consider an uncorrelated NOMO-like
trial function of the absolute coordinates

\begin{equation}
\varphi (x_{1},x_{2},x_{3})=\frac{2^{3/4}\sqrt{b}a^{1/4}}{\pi ^{3/4}}\exp
\left[ -ax_{1}^{2}-b(x_{2}^{2}+x_{3}^{2})\right]  \label{eq:phi_var_2}
\end{equation}
where the optimal values of $a$ and $b$ minimize
$W(a,b)=\left\langle \hat{H}_{d}-\hat{T}_{CM}\right\rangle $. In
this case we have the following TF-NOMO parameters and energy

\begin{eqnarray}
a &=&\frac{\sqrt{3}}{2},\;b=\frac{\sqrt{6(\lambda +1)}}{4}  \nonumber \\
W(a,b) &=&\frac{\sqrt{6(\lambda +1)}}{3}+\frac{\sqrt{3}}{3}
\label{eq:W(a,b)_2}
\end{eqnarray}

In order to measure the effect of keeping the kinetic energy of
the center of mass we also choose the values of the variational
parameters from the minimum of $W(a,b)=\left\langle
\hat{H}_{d}\right\rangle $, which yields
\begin{eqnarray}
a &=&\frac{\sqrt{2}}{2}  \nonumber \\
b &=&\frac{\sqrt{\lambda +1}}{2}  \nonumber \\
W(a,b) &=&\sqrt{\lambda +1}+\frac{\sqrt{2}}{2}  \label{eq:W(a,b)_3}
\end{eqnarray}
From now on we refer to equations (\ref{eq:W(a,b)_2}) and
(\ref{eq:W(a,b)_3}) as TF-NOMO and TC-NOMO in order to make a
connection with the approach of Nakai et al\cite{NHM05}.

\begin{figure}[h]
\begin{center}
\bigskip\bigskip\bigskip \includegraphics[width=9cm]{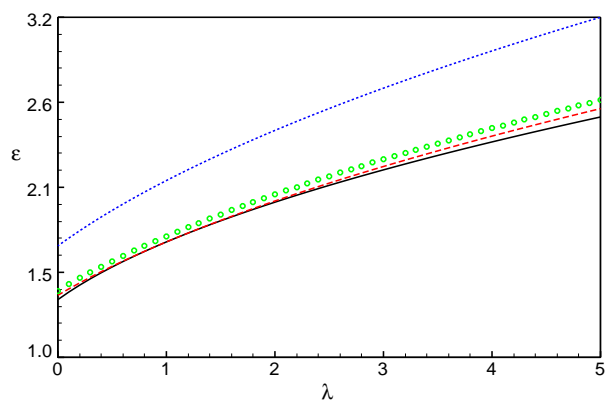}
\end{center}
\caption{Exact (solid line, black), TF-NOMO (dashed line, red),
TC-NOMO (points, blue) and CTC-NOMO (circles, green) ground--state
dimensionless energy} \label{Fig:energy}
\end{figure}

Fig.~\ref{Fig:energy} shows the exact and approximate
dimensionless ground-state energy calculated in the two ways just
outlined. As expected, TF-NOMO yields considerably more accurate
results because TC-NOMO is strongly contaminated with the kinetic
energy of the center of mass. This point was discussed earlier by
Fern\'{a}ndez\cite{F10} with respect to the DEMO
method\cite{TO00}. Note that TF-NOMO is remarkably accurate for
all $\lambda $ and yields the exact result for $\lambda =1$ as
discussed above.

We may try and improve the TC-NOMO results by simply subtracting
the kinetic energy of the center of mass, thus producing a sort of
corrected TC-NOMO or CTC-NOMO:
\begin{equation}
W^{CTCNOMO}=\left\langle \varphi ^{TCNOMO}\right|
\hat{H}_{d}-\hat{T}_{CM}\left| \varphi ^{TCNOMO}\right\rangle
\label{eq:W_CTCNOMO}
\end{equation}
Figure~\ref{Fig:energy} shows that the energy calculated in this
way agrees quite well with the exact and TF-NOMO ones. This result
shows that most of the error in the energy calculated by means of
the TC-NOMO comes from the spurious kinetic energy and just a
relatively small contribution comes from the inadequate
optimization of the variational wavefunction with respect to
$\hat{H}_{d} $.

It is not difficult to explain why the uncorrelated trial function
in absolute coordinates yields such good results (even exact ones
for $\lambda =1$). It we substitute $x_{2}=q_{2}+x_{1}$ and
$x_{3}=q_{3}+x_{1}$ into equation (\ref{eq:phi_var_2}) and
integrate the square of
$\tilde{\varphi}(x_{1},q_{2},q_{3})=\varphi
(x_{1},q_{2}+x_{1},q_{3}+x_{1})$, with respect to $x_{1}$, we
obtain
\begin{equation}
\rho (q_{2},q_{3})=\int_{-\infty }^{\infty }\tilde{\varphi}%
(x_{1},q_{2},q_{3})^{2}\,dx_{1}=\frac{2\sqrt{a}b}{\pi \sqrt{a+2b}}\exp
\left[ -\frac{2c\left( a+b\right) }{a+2b}\left( q_{2}^{2}+q_{3}^{2}\right) +%
\frac{4b^{2}q_{2}q_{3}}{a+2b}\right]
\end{equation}
Note that the use of the absolute coordinates in the trial wavefunction
introduces some sort of correlation between the translation-invariant
coordinates $q_{2}$ and $q_{3}$ when we integrate with respect to the
redundant variable. The resulting correlation is reasonable because it is
determined by the variational method, and, in particular, when $\lambda =1$
it yields (fortuitously) the exact probability distribution for the relative
coordinates

\begin{equation}
\rho (q_{2},q_{3})=\frac{1}{\pi }\exp \left[ -\frac{2}{\sqrt{3}}\left(
q_{2}^{2}+q_{3}^{2}-q_{2}q_{3}\right) \right] =\psi _{00}(q_{2},q_{3})^{2}
\end{equation}
It is now clear why we obtained the exact energy and expectation
values before for this particular case. In fact, we expect to
obtain the exact expectation values of any observable in relative
coordinates. We do not obtain the exact expectation value of
$\hat{T}_{CM}$ because this operator contains a derivative with
respect to the redundant absolute variable that does not appear in
the exact square-integrable wavefunction.

\begin{figure}[h]
\begin{center}
\bigskip\bigskip\bigskip \includegraphics[width=9cm]{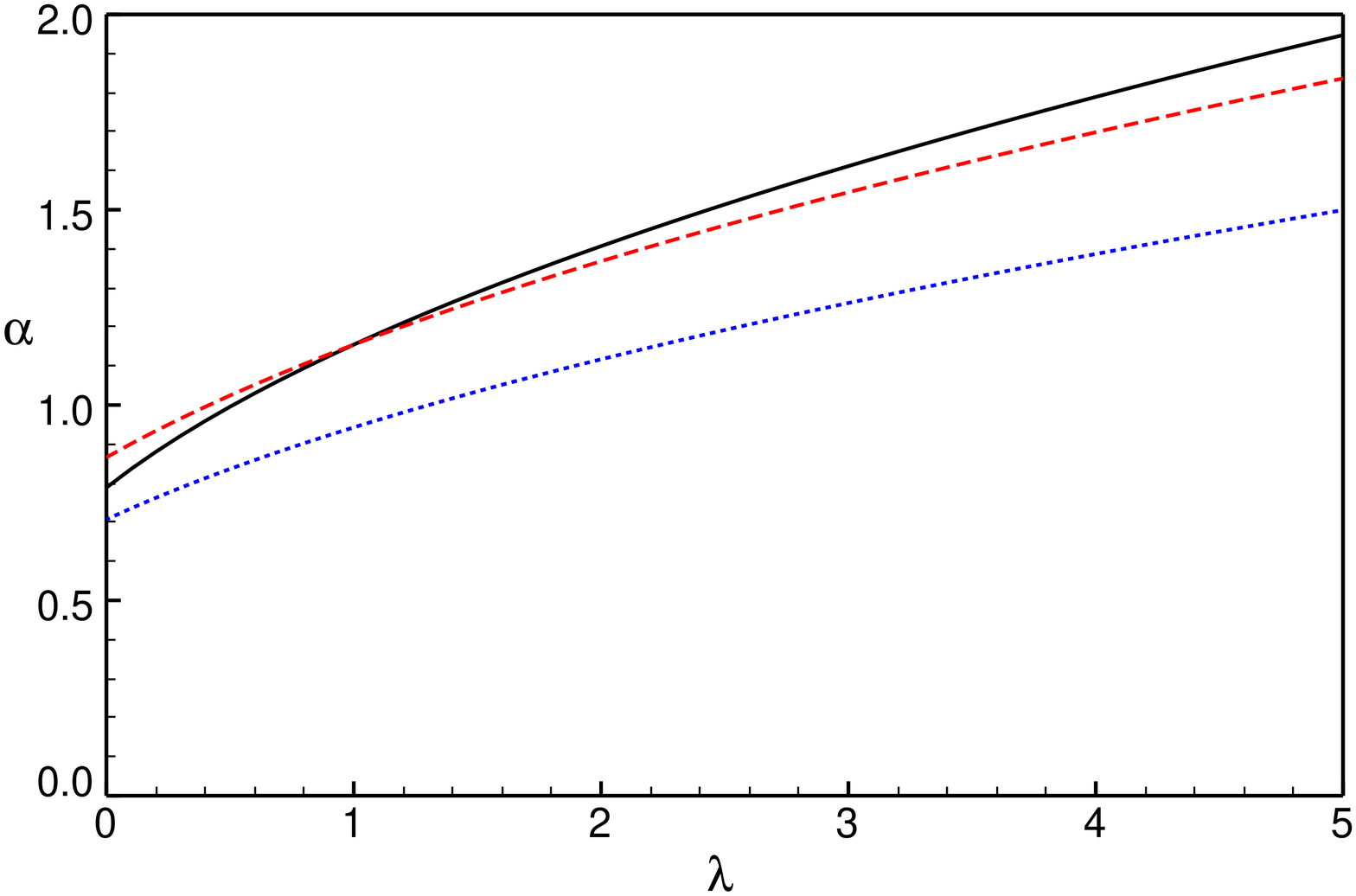}
\end{center}
\caption{Exact (solid line, black), TF-NOMO (dashed line, red) and
TC-NOMO (points, blue) exponential coefficient $\alpha$}
\label{Fig:alpha}
\end{figure}

\begin{figure}[h]
\begin{center}
\bigskip\bigskip\bigskip \includegraphics[width=9cm]{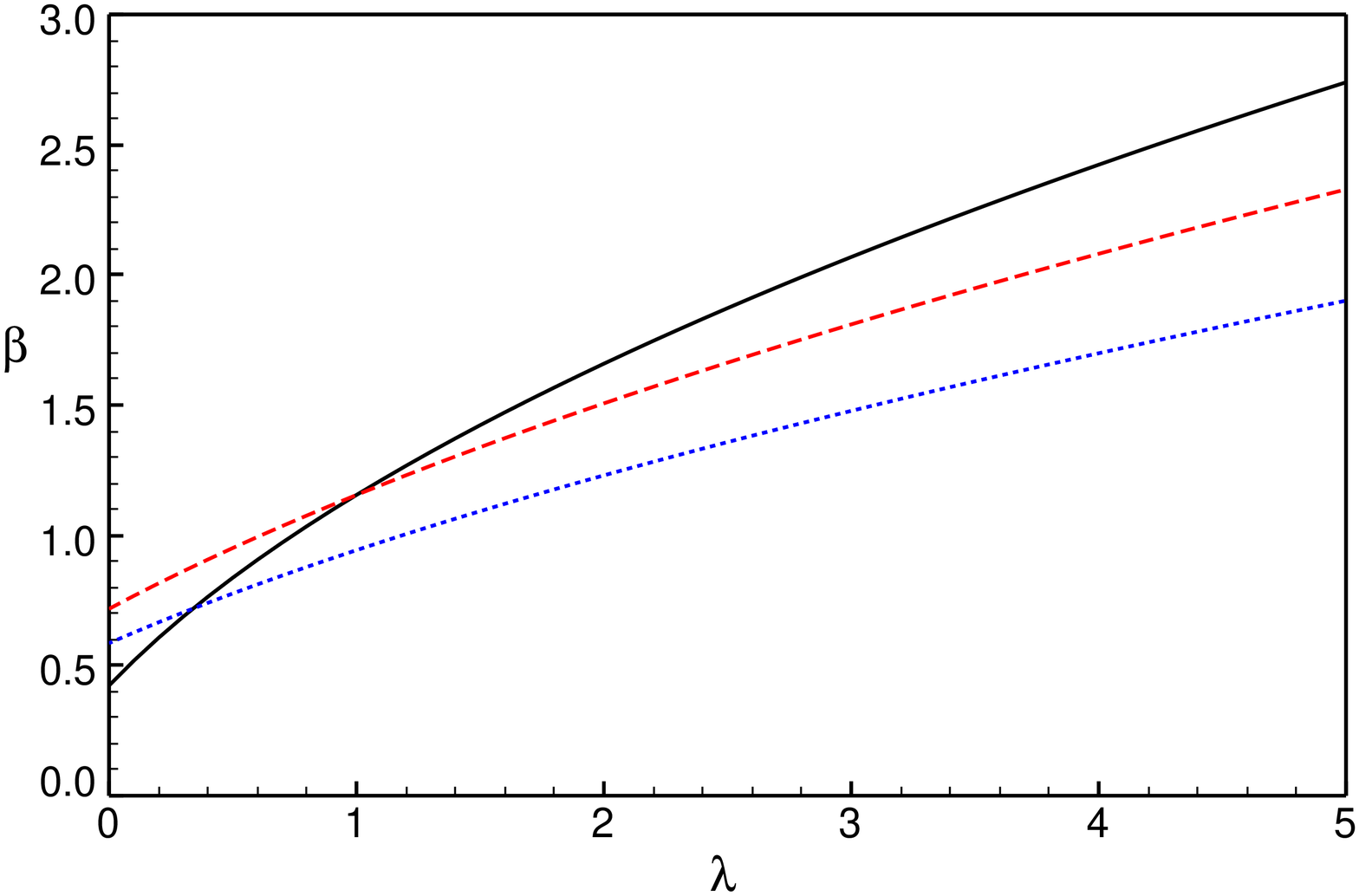}
\end{center}
\caption{Exact (solid line, black), TF-NOMO (dashed line, red) and
TC-NOMO (points, blue) exponential coefficient $\beta$}
\label{Fig:beta}
\end{figure}

In order to compare the approximate and exact wavefunctions we write both $%
\rho (q_{2},q_{3})$ and $\psi _{00}(q_{2},q_{3})^{2}$ as $\left[ \left(
4\alpha ^{2}-\beta ^{2}\right) ^{1/2}/(2\pi )\right] \exp \left[ -\alpha
\left( q_{2}^{2}+q_{3}^{2}\right) +\beta q_{2}q_{3}\right] $ ($2\alpha
>|\beta |$). Figures \ref{Fig:alpha} and \ref{Fig:beta} show that the
exponential coefficients $\alpha $ and $\beta $, respectively,
given by the TF-NOMO agree remarkably well with the exact ones,
whereas the TC-NOMO coefficients are considerably less accurate.
Note that the effect of keeping the kinetic energy of the center
of mass not only affects the energy (which is expected) but also
the form of the variational wavefunction. The agreement between
the exact and TF-NOMO exponential parameters $\alpha$ and $\beta$
explains why the approximate wavefunction yields so accurate
expectation values of operators in the relative coordinates
$q_{2}$ and $q_{3}$. In particular, the TF-NOMO results are exact
for $\lambda =1$, but, of course, such a result is not to be
expected in a realistic case.

If we minimize $\left\langle \varphi \right| \hat{H}_{r}\left| \varphi
\right\rangle $ with an uncorrelated trial function of the relative
coordinates
\begin{equation}
\varphi (q_{2},q_{3})=\sqrt{\frac{2a}{\pi }}\exp \left[
-a(q_{2}^{2}+q_{3}^{2})\right]  \label{eq:phi_var_rel_unc}
\end{equation}
we obtain the optimal variational parameter $a=\sqrt{2\lambda
+2}/4$ and the resulting approximate energy $W(a)=\sqrt{2\lambda
+2}$ is considerably less accurate than that given by Eq.
(\ref{eq:W(a,b)_2}). This result suggests that it is preferable to
apply the NOMO method with orbitals that depend on the absolute
coordinates as long as we remove the kinetic energy of the center
of mass in the optimization process. That is to say, we minimize
the expectation value of $\hat{H}_{r}=\hat{H}-\hat{T}_{CM}$
expressed, for simplicity, in the same set of absolute coordinates
chosen for the NOMO variational wavefunction. Under such
conditions the TF-NOMO-SCF method appears to take into account
part of the correlation energy that in the present simple model is
given by $W(a,b)-W(a)$. This energy difference is quite similar
for TF-NOMO and CTC-NOMO according to Fig.~\ref{Fig:energy}.

\section{Conclusions}

\label{sec:conclusions}

We have tried to elucidate the effect of using absolute
coordinates in the NOMO-SCF variational method. Since such an
analysis for an actual molecule, even as simple as $H_{2}$, is
rather complicated we chose a simple model of three particles with
harmonic interactions in one dimension. Although rather
oversimplified, this model enables us to take into account the
main ingredients of the NOMO-SCF method. We have a
translation-invariant potential-energy function and, consequently,
we have to remove the unbounded motion of the center of mass. In
this case a correlated Gaussian function of the relative
coordinates yields the exact result that is most convenient to
test the approximate ones.

The NOMO-SCF wavefunction is simply a product of Gaussian
functions (orbitals) for each of the particles. The integration of
the square this uncorrelated function with respect to the
redundant absolute coordinate (three in a realistic case as shown
in Eq.~(\ref {eq:rho_gen})) gives rise to some kind of particle
correlation. If the adjustable parameters in this trial function
are optimized variationally with the Hamiltonian
$\hat{H}_{r}=\hat{H}-\hat{T}_{CM}$ then the correlation just
mentioned appears to improve the calculation of the energy and
expectation values considerably. Perhaps, one should not expect
such a remarkable success for an actual molecule, but, however,
there is no doubt that even in that case the NOMO-SCF will take
into account part of the correlation energy in spite of being
based on uncorrelated Gaussian functions. In order to verify this
conjecture that is expressed in Eq.~(\ref{eq:rho_gen}) it is only
necessary to calculate the energy with an uncorrelated NOMO trial
function and $\hat{H}_{r}$ both in terms of relative coordinates
(like present Eq.~(\ref{eq:phi_var_rel_unc})). Note that this
interesting feature of the TF-NOMO has apparently passed unnoticed
in the applications of the method\cite{NHM05, HN06,MHN06,N07}.
This fact reinforces our claim on the utility of simple models for
the study of rather complicated problems.

The simple model also shows that if we simply subtract the
expectation value of the kinetic energy of the center of mass then
the resulting energy is quite accurate (what we have called
CTC-NOMO). However, in order to improve the calculation of the
expectation values of other observables it is convenient to apply
the SCF procedure with the relative Hamiltonian operator
$\hat{H}_{r}=\hat{H}-\hat{T}_{CM}$ that leads to what is commonly
known as TF-NOMO\cite {NHM05, HN06,MHN06,N07}.

\end{document}